\documentclass{sig-alternate}

\usepackage{acronym}
\usepackage{subfigure}
\usepackage{booktabs}
\usepackage{algorithm}
\usepackage{algcompatible}
\usepackage{url}
\usepackage{color}
\usepackage{graphicx}

\makeatletter
\def\@copyrightspace{\relax}
\makeatother

\begin{document}
%

\title{Time Warp on the Go (Updated Version)\footnotemark}

\def\sharedaffiliation{%
\end{tabular}
\begin{tabular}{c}}

\numberofauthors{3}
\author{
\alignauthor Gabriele D'Angelo\\
\email{g.dangelo@unibo.it}
\alignauthor Stefano Ferretti\\
\email{sferrett@cs.unibo.it}
\alignauthor Moreno Marzolla\\
\email{marzolla@cs.unibo.it}
\sharedaffiliation
\affaddr{Department of Computer Science}\\
\affaddr{University of Bologna}\\
\affaddr{Mura A. Zamboni 7, I-40127 Bologna, Italy}\\
}

\maketitle

\footnotetext{The publisher version of this paper is available at \url{http://dx.doi.org/10.4108/icst.simutools.2012.247736}. \textbf{{\color{red}Please cite as: Gabriele D'Angelo, Stefano Ferretti, Moreno Marzolla. Time Warp on the Go. Proceedings of 3nd ICST/CREATE-NET Workshop on DIstributed SImulation and Online gaming (DISIO 2012). In conjunction with SIMUTools 2012. Desenzano, Italy, March 2012. ISBN: 978-1-936968-47-3.}}}

\abstract{In this paper we deal with the impact of multi and many-core
  processor architectures on simulation. Despite the fact that modern
  CPUs have an increasingly large number of cores, most softwares are
  still unable to take advantage of them. In the last years, many
  tools, programming languages and general methodologies have been
  proposed to help building scalable applications for multi-core
  architectures, but those solutions are somewhat limited. Parallel
  and distributed simulation is an interesting application area in
  which efficient and scalable multi-core implementations would be
  desirable. In this paper we investigate the use of the Go
  Programming Language to implement optimistic parallel simulations
  based on the Time Warp mechanism. Specifically, we describe the
  design, implementation and evaluation of a new parallel simulator.
  The scalability of the simulator is studied when in presence of 
  a modern multi-core CPU and the effects of the Hyper-Threading 
  technology on optimistic simulation are analyzed.}


\section{Introduction}

A recent trend in computing is the availability of CPUs with more and
more execution cores. In a multi-core processor, two or more
independent execution units (cores) are packaged in the same die.
This means that multi-core CPUs are actually shared-memory, Multiple
Instructions, Multiple Data (MIMD) machines capable of running
multiple independent instructions at the same time. The first
generation of multi-core CPUs was equipped with two cores only, but
currently available processors have four or more cores. For some
specific fields, processors with one hundred cores are already on the
market~\cite{tilera100}. With such premises, it is clear that in the
next years multi-core processors will be replaced by many-core
processing architectures.\\

This trend has a strong impact on software, since sequential
algorithms are unable to efficiently exploit the computational power
provided by modern CPUs. This is true not only for servers and High
Performance Computing (HPC) environments but also for desktop PCs. It
is well known that developing parallel algorithms and implementing
them is much harder than sequential ones. To address this issue, many
programming languages and tools have been proposed, including:
OpenMP~\cite{openMP}, CUDA~\cite{CUDA}, OpenCL~\cite{openCL}, Intel
Threading Building Blocks (TBB)~\cite{TBB}, Go Programming
Language~\cite{Go}, Erlang~\cite{Erlang}, MapReduce~\cite{MapReduce}
(just to name a few). Each of them comes with its peculiarities and
specific field of application. Despite a quite large research effort
in finding new paradigms to tackle these new computation
architectures, a large consensus has still not being reached. In many
fields, designers and developers are still trying to understand what
is the impact of these technologies and what is the more promising one
for their needs.\\

If we restrict our focus on the field of discrete-event simulation,
things are not better. For many reasons~\cite{gda-hpcs-11}, most
simulators are still based on sequential approaches and therefore
unable to take advantage of more than one CPU core. Given the trend
towards many-core processors described above, to improve the execution
speed of simulators it is no longer possible to count on the increase
of CPU clock speed. In other words, to take full advantage of new
multi (many) core architectures it is necessary to use Parallel And
Distributed Simulation (PADS)~\cite{Fujimoto:1989:PDE:76738.76741}
techniques, even for running a simulation on a normal desktop
PC. Parallel simulation techniques that were used by niche users only,
now need to go mainstream. However, in order to gain any support from 
simulation model developers, multi-core PADS libraries and tools should 
hide low-level details and present a convenient interface to users.
While it is safe to assume that~PADS middlewares are designed and 
implemented by experts in the field, it is important not to assume 
that users of these tools are~PADS experts as well. It is worth noting
that these simulation tools will not be sufficient for exploiting 
the resources provided by the many-cores architectures that will be 
available in the next years. Also the simulation models will have to 
be built in a way that allows their parallel execution.\\

In this paper we describe the design and implementation of a new
parallel simulation tool (called Go-Warp), based on the Go Programming
Language~\cite{Go}. The Go language has good support for concurrency
and communication; many features introduced in Go seem to have a good
potential but its usage in PADS is still unexplored.\\

The rest of this paper is organized as
follows. Section~\ref{sec:background} provides a background on
parallel and distributes simulation issues. In
Section~\ref{sec:related} we review the relevant
literature. Section~\ref{sec:go} briefly outlines the most interesting
aspects of the Go Programming Language. Section~\ref{sec:go-warp}
introduces the Go-Warp simulator that we have designed and implemented
and Section~\ref{sec:go-performance} shows an initial performance
evaluation of this tool. Finally, Section~\ref{sec:conc} provides some
concluding remarks.

\section{Background}\label{sec:background}

Many different approaches have been proposed for building simulators,
Discrete Event Simulation~\cite{Law:1999:SMA:554952} is one of the
more popular. In this approach the system under study is modeled
through a set of state variables. Each update in the simulated system (called
event) happens at a discrete time instant and is reflected to the
state variables. This means that the evolution of the simulation is
obtained through the creation, delivery and execution of a sequence of
events, ordered according to their timestamps (time of occurrence). In
a sequential (monolithic) simulator a single CPU (execution unit) is
responsible for executing all the events in the correct
order. Sequential execution of events obviously limits the scalability
of the simulator, which in turn limits the complexity of models which
can be executed.

In Parallel Discrete Event Simulation
(PDES)~\cite{Fujimoto:1989:PDE:76738.76741} a set of execution units
(e.g.~CPUs, cores or hosts) runs the simulation. In this case, the
simulated model is partitioned among the execution units. While this
improves scalability, it also introduces communication and
synchronization issues as each execution unit produces events that may
be be delivered to other units. To ensure that causality is not
violated (that is, to ensure that events are processed in the correct
order), each execution unit must be synchronized with the others. It
is worth noting that a simulation in which causal dependencies among
events are violated can not be considered correct and produces results
that have no validity.

\subsection{Parallel and Distributed Simulation}
\label{sec:sub:background-PADS}

The PDES approach described in the previous section can be implemented
using a Parallel And Distributed Simulation (PADS). A PADS is obtained
through the interconnection of a set of model components, usually
called Logical Processes (LPs). Each LP is responsible for a part of
the system and needs to be coordinated with other LPs for
synchronization and data distribution~\cite{FUJ00}. What happens is
that each LP is usually executed by a processor (or a processor
core). The basic difference between a sequential simulation and a PADS
is the lack of a global state, that is a global vision on the
simulated model state and its evolution. The distributed nature of the
system and the presence of a network that interconnects the different
parts of the execution architecture has some important consequences:

\begin{itemize}

\item the simulated model has to be partitioned among the
  LPs~\cite{bagrodia98}. Increasing the number of execution units
  means a higher number of parts. In some cases, this
  \textbf{partitioning} is simplified by the nature of the simulated
  system, e.g., when the system under test can ``naturally'' be
  described as a set of interacting objects. In other cases,
  partitioning is much harder, e.g., when the system is intrinsically
  monolithic. In any case, given the parallel/distributed nature of
  the simulator there are some extra factors to take into
  consideration when partitioning the model, such as minimizing the
  amount of network communication between LPs and balancing the
  workload among the execution units;

\item the simulation traces obtained by the PADS have to be identical
  to the ones that would have been obtained using a sequential
  simulator. Clearly, this is possible only if
  \textbf{synchronization} mechanisms are implemented. In other words,
  the execution of each LP has to be properly synchronized;

\item each LP will produce data (i.e.~state updates) that are relevant
  for other LPs. For performance reasons, this \textbf{data
    distribution} cannot be implemented using broadcast. The correct
  approach is to match the data production with the expression of
  interest and therefore delivering only the necessary
  data~\cite{Jun:2002:ESM:564062.564074}.

\end{itemize}

All these aspects deserve attention but synchronization remains very
relevant. This because it has a very deep impact on simulator
performance and because the choice of the synchronization mechanism
shapes the design and implementation of the simulation model.

\subsection{Synchronization}
\label{sec:sub:background-synchro}

The correct implementation of a PADS requires that all events are
timestamped, encapsulated in a message and delivered. Following
Lamport's definition~\cite{Lamport1978}, two events are said to be in
causal relation if one of them can have some consequences on the
other. Causal relations induces a partial order among the events:
breaking this order produces a causality violation, which means that
the simulator is incorrectly evaluating the model. Avoiding causality
violations is easy in sequential (i.e.,~monolithic) simulators: all
events must be considered in non-decreasing timestamp
order. Unfortunately, the problem is much harder in a
parallel/distributed setting, since every execution unit can simulate
its portion of the model at a different speed and the interconnection
network can introduce unpredictable delays, jitter, and packet losses.
Therefore, in this case processing the events in causal order requires
that all the LPs in the PADS are coordinated using a synchronization
algorithm.

In the past, a lot of work has been done in this field. Many different
synchronization algorithms have been proposed, which can be grouped in
three main approaches:

\begin{itemize}

\item \emph{time-stepped}: the simulated time is advanced according to
  fixed-size time steps. This means that before proceeding to the next 
  time\-step each LP has to wait that all other LPs in the simulation 
  have finished the current time\-step~\cite{1261535}. The design and 
  implementation of this approach is quite simple, but can be 
  inappropriate for some simulation models. For example if the system 
  is hard to model in time-steps or if the size of the time-steps
  needs to be too small;

\item \emph{conservative}: the goal of this approach is to prevent
  causality violations. This means that, before processing an event
  with timestamp $t$, the LP has to decide if this event is ``safe''
  or not. The event is ``safe'' if, in the future, there will be no
  other events with timestamp less than $t$. It is easy to demonstrate
  that, if this constraint is followed by all LPs, there will be no
  causality violations. Many algorithms can be used to guarantee that
  this constraint is not violated; the Chandy-Misra-Bryant (CMB)
  approach~\cite{misra86} is the most used. More in detail, the CMB
  approach is based on three main assumptions:

  (\emph{i}) each LP has as many incoming queues as LPs from which it can receive events;
  (\emph{ii}) all the generated events produced by the local LP must be sent out in non decreasing order;
  (\emph{iii}) the communication between the LPs is reliable and messages are delivered in order.
 
  Before processing an event, each LP must check all incoming queues
  to find what is the next safe event. If there are no empty queues,
  then the incoming event with lowest timestamp is safe and can be
  processed. If there are empty queues, the LP must wait for at least
  one event to appear on them. Obviously, this mechanism is deadlock
  prone; to avoid deadlocks a new type of message (called NULL
  message) is introduced. If an LP $X$ sends a NULL message to LP $Y$
  with a given timestamp $t$, then $X$ is telling $Y$ that it will not
  even send any proper message with timestamp less than $t$. This
  allows the receiving LP $Y$ to properly compute the next safe time
  and advance the simulation. The main drawbacks of this approach is
  that NULL messages increase the communication traffic, and also that
  deciding if and when a NULL message can be generated requires
  knowledge of the simulation model;

\item \emph{optimistic}: in this case the LPs are free to process the
  events in receiving order without any check for safety. In other
  words, the LP does nothing to avoid causality violations. Of course,
  many factors such as CPU speed, network load and model complexity
  can delay the arrival of messages. Late events, usually called
  straggler messages, lead to causality violations. The arrival of an
  event with a lower timestamp than the current local simulated time
  represents a problem for the receiving LP. To fix the problem, the
  receiving LP executes a roll-back of all model state variables to a
  previous version that is considered correct. It is worth noting that
  the roll-back has to be propagated to all other affected
  LPs~\cite{timewarp}. This can lead to a cascade of rollbacks that
  brings the simulator back to a previous state. To be able to perform
  a roll-back when necessary, each LP must keep some information such
  as all changes to the state variables and all sent events. Such
  information can require a lot of memory, which can be periodically
  reclaimed (fossil collection~\cite{FUJ00}) by computing a Global
  Virtual Time (GVT)~\cite{samadi} which represents a safe lower bound
  on the global simulation time. In other words, no events with
  timestamp prior to the GVT will ever appear in the future, and so
  each LP can reclaim all state variables preceding the GVT.
  
\end{itemize}

\section{Related works}
\label{sec:related}

Over the years a large amount of research work has been done for
improving the performance of optimistic synchronization algorithms and
for adapting them to different execution environments. The Time Warp
algorithm was used for building simulations on top of clusters with
hundreds of thousands GPUs~\cite{10.1109/DS-RT.2009.41} with very good
performances. In~\cite{nianle09} the authors describe an optimistic
simulator which uses MPI~\cite{usingmpi} as the communication library;
the simulator is evaluated on a multi-core processor under the Windows
OS. It should be observed that MPI, being originally developed for
communication over a LAN, introduces a significant overhead which can
be avoided by allowing LPs to communicate using the shared memory.

Some work has been done for improving the execution speed of Time Warp
on many-core architectures~\cite{simul2011,5936752} but little work
has been done for extending the parallelization up to the simulation
models~\cite{Himmelspach:2010:ESS:1935937.1936120} and to implement an
approach that is more tailored for these new architectures.

\section{The Go Programming Language}
\label{sec:go}

Go is a general purpose programming language announced by Google in
the late 2009 and now developed as an Open Source
project~\cite{Go}. The main goals of this effort is to build a
language that is easy, clean and efficient. The compilation process is
designed to be easier and faster than in traditional languages. Go
provides mechanisms for concurrent execution and inter-process
communication, which facilitate the development of parallel
applications. All these mechanisms are part of the language core and
not provided as external libraries. Finally, Go uses a modern garbage
collector which relieve the programmer from the burden of dynamic
memory management.\\

The main language construct introduced by Go for concurrent
programming is the \emph{goroutine}. A goroutine is a function
executing in parallel with other goroutines in the same address space.
Goroutines can communicate through shared memory; furthermore, they
are implemented using a lightweight approach, so they introduce a low
overhead. Specifically, goroutines are multiplexed onto multiple
operating system threads, meaning that if a goroutine is blocked
waiting for some input, other goroutines can continue to
run. Recently, a new feature added to the language permits to pack
multiple goroutines in the same thread and therefore reduce the
overhead in programs using hundreds of goroutines. It is worth noting
that implementing a goroutine is very easy: the programmer has to
prefix a function or method call with the ``go'' keyword. In this way,
all the complexities of thread management are transparent for the
programmer.

The communication between goroutines is implemented using another
interesting language construct, the so called \emph{chan} (that stands
for \emph{channel}). A chan is a data type that can be used for both
communication and synchronization between goroutines. Each chan has a
capacity that is the size of the buffer in the channel. If the
capacity of a channel is zero then the channel is synchronous and can
be used only for synchronization. In all other cases, the channel is
asynchronous and can be used for the transmission of typed messages.\\

The Go project is under active development: the main design is
complete but many implementation aspects are yet to be finished. In
particular, as stated in the official website: ``one of Go's design
goals is to approach the performance of C for comparable programs, yet
on some benchmarks it does quite poorly''~\cite{Go}. The internal
scheduler that manages the goroutines is among the parts that is far
from being finished. In the current version, the runtime is unable to
automatically determine what is the maximum number of CPUs (cores)
that can be executing simultaneously. This means that the programmer
has to set this parameter using the GOMAXPROCS function. It is
expected that this need will go away when the scheduler improves in
future versions.

\section{Design and Implementation of Go-Warp}
\label{sec:go-warp}

Go-Warp is a simulator based on the Time Warp synchronization
algorithm; Time Warp was originally proposed by
Jefferson~\cite{timewarp}, and briefly introduced in
Section~\ref{sec:background}. Go-Warp uses Samadi's
algorithm~\cite{samadi} for the calculation of the Global Virtual Time
(GVT). This algorithm is quite simple to implement but adequate to
perform a preliminary performance evaluation of the simulator. In the
next versions, more complex GVT algorithms such as
Mattern's~\cite{Mattern:1993:EAD:167323.167331} will be added.\\

A Go-Warp simulation is composed of one or more LPs, each one being
executed independently from others and implemented using a
goroutine. Following the design suggested by the Go language, the
LP-to-LP communication is realized through asynchronous chans. This
means that the LPs are able to send data without blocking and proceed
with the execution. Each LP accesses some shared variables that used
for the efficient implementation of some functionalities (such as the
GVT calculation). As said before, following the optimistic approach
there is no ``a priori'' attempt to synchronize the LPs, the access to
global variables is controlled only by a mutex implemented using the
language primitives.\\

The design of Go-Warp is based on the tasks required by the Time Warp
algorithm. First of all, the LP has to receive, deliver, store and
process events that are encapsulated in messages. Moreover, it has to
handle the rollbacks caused by straggler messages. Finally, it has to
run its local part of the GVT calculation algorithm. Events management
is the core activity: in Go-Warp, each LP uses a priority queue
implemented using a min-heap (called GoHeap) to store the future
events that have to be processed. For performance reasons, each GoHeap
node is an array that contains all the local events with the same
timestamp. Every LP has a GoLocalState structure that contains all the
LP local data: the current simulated time, the LP unique identifier, a
GoHeap instance for storing the events to be processed, a list for
maintaining the processed events that will be needed in case of
rollbacks, a list for messages sent to other LPs (used to propagate
rollbacks, if necessary), a list of rollback requests from other LPs
(i.e.,~the so called anti-messages), and a list for storing the
messages that have been sent by the local LP but that are still not
acknowledged by the recipient (this is needed by the Samadi's GVT
computation algorithm).\\

By now, in Go-Warp each LP is implemented using a single goroutine,
meaning that all activities performed by an LP are executed by a
single thread. We are working on enhancing the internal parallelism 
of the LPs, which will potentially improve performance in many-cores 
CPUs. As said before, future CPUs will have a very large number of 
cores. You could try to exploit them with a very aggressive 
partitioning of the simulation model (i.e.~using a high number of 
LPs), but to obtain a good partitioning is not an easy task (see 
Section~\ref{sec:sub:background-PADS}) and it becomes even 
harder with the addition of more LPs. Otherwise, you could use a 
limited number of partitions and work on the LP internals, for 
example with the parallel execution of some of the LP mechanisms 
(e.g.~synchronization, data distribution) and with simulation models
that can be implemented with a better degree of parallelism.
It is worth noting that, in the current version of the simulator, 
the number of CPU cores used at runtime is set manually using the 
GOMAXPROCS function (described in the previous section). The
goal is to run each LP in a different core to minimize context
switches, that are usually quite costly in terms of overhead. This
approach is clearly not optimal, it will require some future work and
a better support from the Go runtime scheduler.\\

We plan to make the Go-Warp simulator freely available, in both binary
and source form, on the research group website~\cite{pads}. The software
distribution will include all the tools, configurations and models 
used to conduct the performance evaluation shown in this paper.

\section{Performance Evaluation}
\label{sec:go-performance}

To evaluate the performance of Go-Warp we used a synthetic benchmark
called PHOLD~\cite{FujimotoPHOLD}, that is a model specifically
designed for the performance evaluation of Time Warp
implementations. PHOLD is the parallel version of the HOLD benchmark
for event queues~\cite{Jones:1986:ECP:5684.5686} and it is quite
simple to implement. Each PHOLD model is made by a set of entities
that are partitioned among the LPs; each LP contains the same number
of entities. Each entity in the simulation produces and consumes
events. When an entity consumes an event, a new event is generated and delivered to
another entity (note that the total number of events in the system
remains constant). The timestamp of the new event is computed by
adding an exponentially distributed random number with mean $5.0$ to the
timestamp of the receiving event. In our implementation the recipient
is randomly chosen using a uniform distribution. When a LP processes an event, 
a new event is generated and delivered to another entity in the simulation. 
In our implementation the recipient entity is randomly chosen using a uniform
distribution. In this way, the total number of events in the system is
fixed and the model is almost in steady state~\cite{Ewald}.\\

There are four main parameters which are used to control the
benchmark. The first one is the number of LPs and the second is 
the number of entities that are simulated. The third parameter 
(called event density) is defined as the percentage of entities 
starting the simulation generating an event.

The forth and last model parameter is the workload, that is the 
amount of synthetic work that is executed by each LP every time an 
event is processed. In our case, we implemented the workload as 
a pre-defined number of floating point operations. Using the 
parameters above it is possible to fine tune the
PHOLD model. Increasing the number of simulated entities has the
effect of adding more computation (workload) and communication
(events) to the benchmark. Changing the event density permits to
obtain a model that is more communication bounded and, conversely,
adding more workload results in a computation bounded model.\\

The results shown in this section have been collected using an
Intel(R) Core(TM) i7-2600 CPU \@ 3.40GHz with 4 cores and
Hyper-Threading (HT) technology. The PC\footnote{Obtained by one of
  the authors using personal savings.} has 8 GB of RAM and runs Ubuntu
11.10 (x86\_64 GNU/Linux, 3.0.0-15-generic \#26-Ubuntu SMP).  To
produce statistically valid results, we performed multiple runs for
each experiment and the average values are shown. HT is a technology
introduced by Intel in some of its CPUs for supporting simultaneous
multi-threading~\cite{HT}. HT works by duplicating some parts of the
processor except the the main execution units. From the point of view
of the Operating System, each physical processor core corresponds to
two ``virtual'' processors. This means that the Intel Core i7 CPU used
in this study has 4 physical cores that are seen as 8 virtual
processors by the OS and the applications. The effect of HT on
parallel simulation is not widely studied~\cite{gda-dsrt-2006}, in
particular, the impact of virtual cores on the performance of
optimistic simulation need to be investigated more in deep.\\

The performance of Go-Warp are analyzed by running the PHOLD model for
$1000$ time units, using a fixed number of entities ($1500$), an event
density of $50\%$ and a workload of $10000$ fixed point operations (FPops) 
per simulation event. In Table~\ref{table:wct}
we show the average physical wall-clock time needed to complete a
simulation run, as a function of the number of LPs in which the model
is partitioned, and as a function of the number of (virtual) processor
cores used. Obviously, using more LPs than cores is not a 
good choice for at least two reasons. Firstly, because the 
overhead induced by the context switches could be quite high.
Secondly, because it has been widely demonstrated that Time Warp obtains
good performance only if all the LPs can run at about the same speed.
Having more LPs than cores would introduce imbalances in the system 
and therefore a higher number of straggler messages (and of rollbacks).
As said above, the i7 CPU used in the test has 4 physical cores that
are seen as 8 virtual processors. From the left part of the table
(1--4 cores) it is clear that the best choice is to have as many LPs
as available physical cores. If we consider the right part of the
table (5--8 ``virtual'' cores), the situation is slightly different
due to HT.  In this latter case, using a number of LPs equal to the
number of ``virtual'' cores is not optimal, and the best result is
obtained when running 5 LPs on 6 cores. Table~\ref{table:speed} shows
the speedup obtained in this experiment and confirms the effect of
HT. The speedup is the ratio of the execution time of the sequential
algorithm ($LP=1$) and the execution time of the parallel version with
$n$ LPs. The best speedup is obtained with 5 LPs and 7 cores, this
means that HT is capable of a little increase in the performance.

\begin{table}[t]
\centering%
\begin{scriptsize}
\noindent\makebox[\columnwidth]{
\begin{tabular}{crrrr|rrrr}
\toprule
& \multicolumn{8}{c}{\bf Number of Cores}\\
\cmidrule{2-9}
{\bf \#LPs} & 1 & 2 & 3 & 4 & 5 & 6 & 7 & 8\\
\midrule
1 & \textbf{1704} & 1691 & 1701 & 1685 & 1700 & 1683 & 1703 & 1685 \\
2 &      & \textbf{1050} & 1049 & 1051 & 1056 & 1047 & 1049 & 1050 \\  
3 &      &      & \textbf{864}  & 854  & 858  & 856  & 865  & 853 \\
4 &      &      &      & \textbf{787}  & 799  & 787  & 807  & \textbf{785} \\
5 &      &      &      &      & \textbf{795}  & \textbf{\textcolor{red}{775}}  & \textbf{778}  & 790 \\
6 &      &      &      &      &      & 817  & 823  & 822 \\
7 &      &      &      &      &      &      & 817  & 842 \\
8 &      &      &      &      &      &      &      & 908 \\
\bottomrule
\end{tabular}}
\end{scriptsize}
\caption{Average Wall Clock Time of the simulation run (in milliseconds)}\label{table:wct}
\end{table}

\begin{table}[t]
\centering%
\begin{scriptsize}
\begin{tabular}{crrrr|rrrr}
\toprule
& \multicolumn{8}{c}{\bf Number of Cores}\\
\cmidrule{2-9}
{\bf \#LPs} & 1 & 2 & 3 & 4 & 5 & 6 & 7 & 8\\
\midrule
1 & \textbf{1} & 1 & 1 & 1 & 1 & 1 & 1 & 1 \\
2 &      & \textbf{1.61} & 1.62 & 1.60 & 1.61 & 1.61 & 1.62 & 1.60 \\  
3 &      &      & \textbf{1.97}  & 1.97  & 1.98  & 1.97  & 1.97  & 1.98 \\
4 &      &      &      & \textbf{2.14}  & 2.13  & 2.14  & 2.11  & \textbf{2.15} \\
5 &      &      &      &      & \textbf{2.14}  & \textbf{2.17}  & \textbf{\textcolor{red}{2.19}} & 2.13 \\
6 &      &      &      &      &      & 2.06  & 2.07  & 2.05 \\
7 &      &      &      &      &      &      & 2.08  & 2.00 \\
8 &      &      &      &      &      &      &      & 1.86 \\
\bottomrule
\end{tabular}
\end{scriptsize}
\caption{Speedup with increasing number of cores}\label{table:speed}
\end{table}

It is known that, in general, the performance of a simulator are
strongly influenced by the model characteristics. In
Table~\ref{table:speedentities} we show the speedup obtained with
increasing number of entities (the data are also shown in
Figure~\ref{fig:speedup}). The model with $1000$ entities (green line)
is communication bound and therefore increasing the number of LPs (and
cores) does not yield significant improvements. When using more than 4
LPs, the communication overhead is so high that the speedup actually
decreases. When $6000$ and $11000$ entities are simulated (violet and
pink lines, respectively) the computation load is higher and
increasing the number of LPs gives a slightly better speedup. In all
these cases, the scalability is quite good up to 4 LPs but for larger
values the results deviate from the optimal speedup due to HT.

\begin{table*}[t]
\centering
\begin{tabular}{crrrrrrrrrrr} 
\toprule
& \multicolumn{11}{c}{\bf Number of entities}\\
\cmidrule{2-12}
{\bf \#LPs} & 1000 & 2000 & 3000 & 4000 & 5000 & 6000 & 7000 & 8000 & 9000 & 10000 & 11000 \\ [0.5ex] 
\midrule
1 & 1 & 1 & 1 & 1 & 1 & 1 & 1 & 1 & 1 & 1 & 1 \\ 
2 & 1.58 & 1.65 & 1.68 & 1.68 & 1.69 & 1.71 & 1.72 & 1.74 & 1.73 & 1.74 & 1.74 \\  
3 & 1.81 & 2.05 & 2.17 & 2.19 & 2.19 & 2.24 & 2.25 & 2.29 & 2.27 & 2.28 & 2.34 \\
4 & \textbf{1.87} & 2.33 & 2.40 & 2.53 & 2.55 & 2.64 & 2.65 & 2.72 & 2.66 & 2.72 & 2.76 \\ 
5 & 1.12 & & & & & 2.83 & & & & & 2.96 \\
6 & 0.94 & & & & & 3.01 & & & & & 3.28 \\
7 & 0.80 & & & & & 3.19 & & & & & 3.38 \\
8 & 0.79 & & & & & \textbf{3.23} & & & & & \textbf{\textcolor{red}{3.69}} \\ 
\bottomrule
\end{tabular}
\caption{Speedup, increasing number of simulated entities}\label{table:speedentities} 
\end{table*}

\begin{figure}
\centering
\includegraphics[width=6.0cm,angle=270]{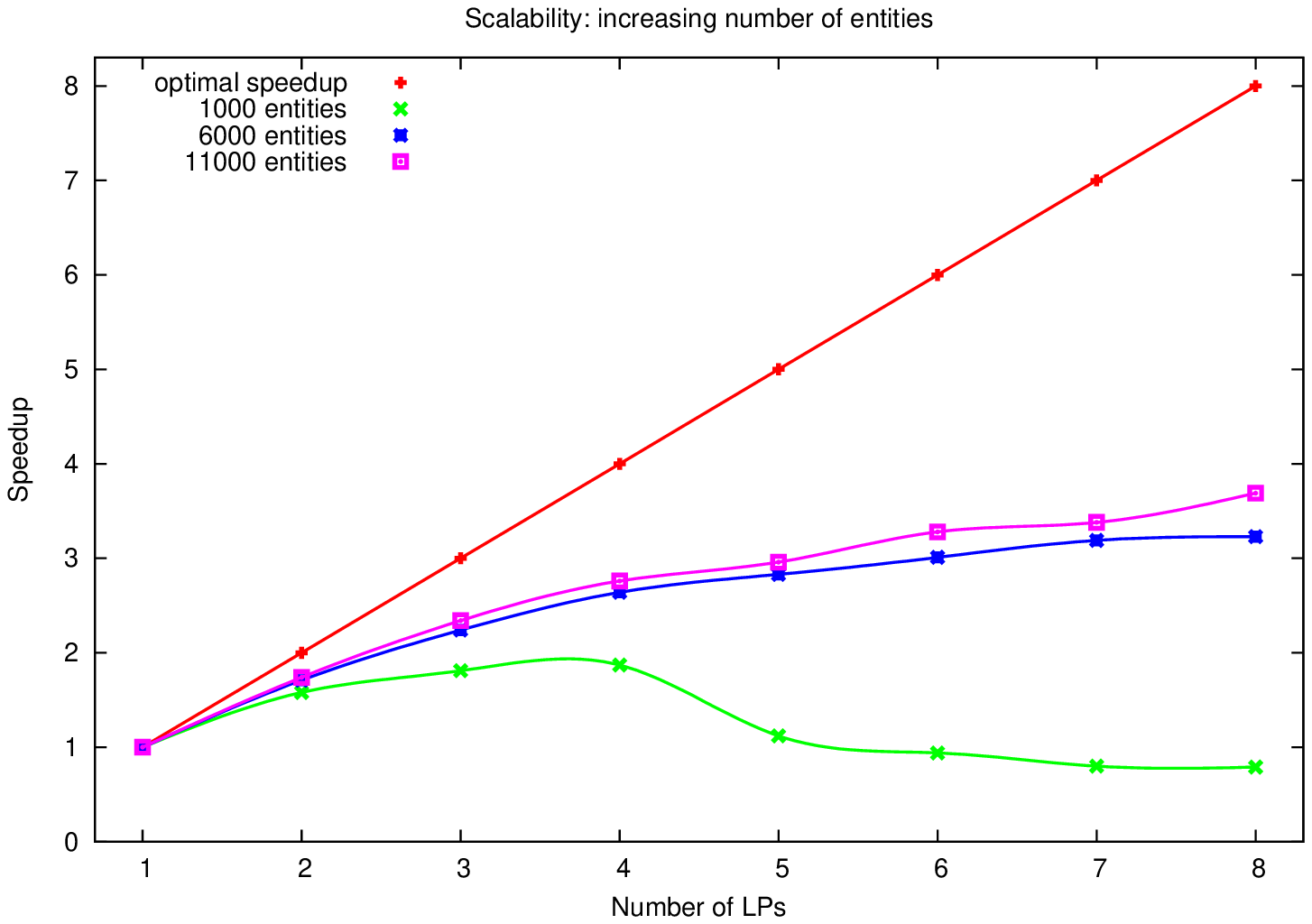}
\caption{Speedup, increasing number of simulated entities, different configurations}
\label{fig:speedup}
\end{figure}

In the last experiment we have simulated a medium number of entities
($6000$) varying the workload. In Figure~\ref{fig:workload} we show
the speedup with a workload of $1000$ (green line), $10000$ (blue
line) and $100000$ (pink line) FPops per event. Increasing the
workload produces a benchmark that is more computation bound. As
expected, increasing the FPops gives very good speedup results that
are near to the theoretical limit.

\begin{figure}
\centering
\includegraphics[width=6.0cm,angle=270]{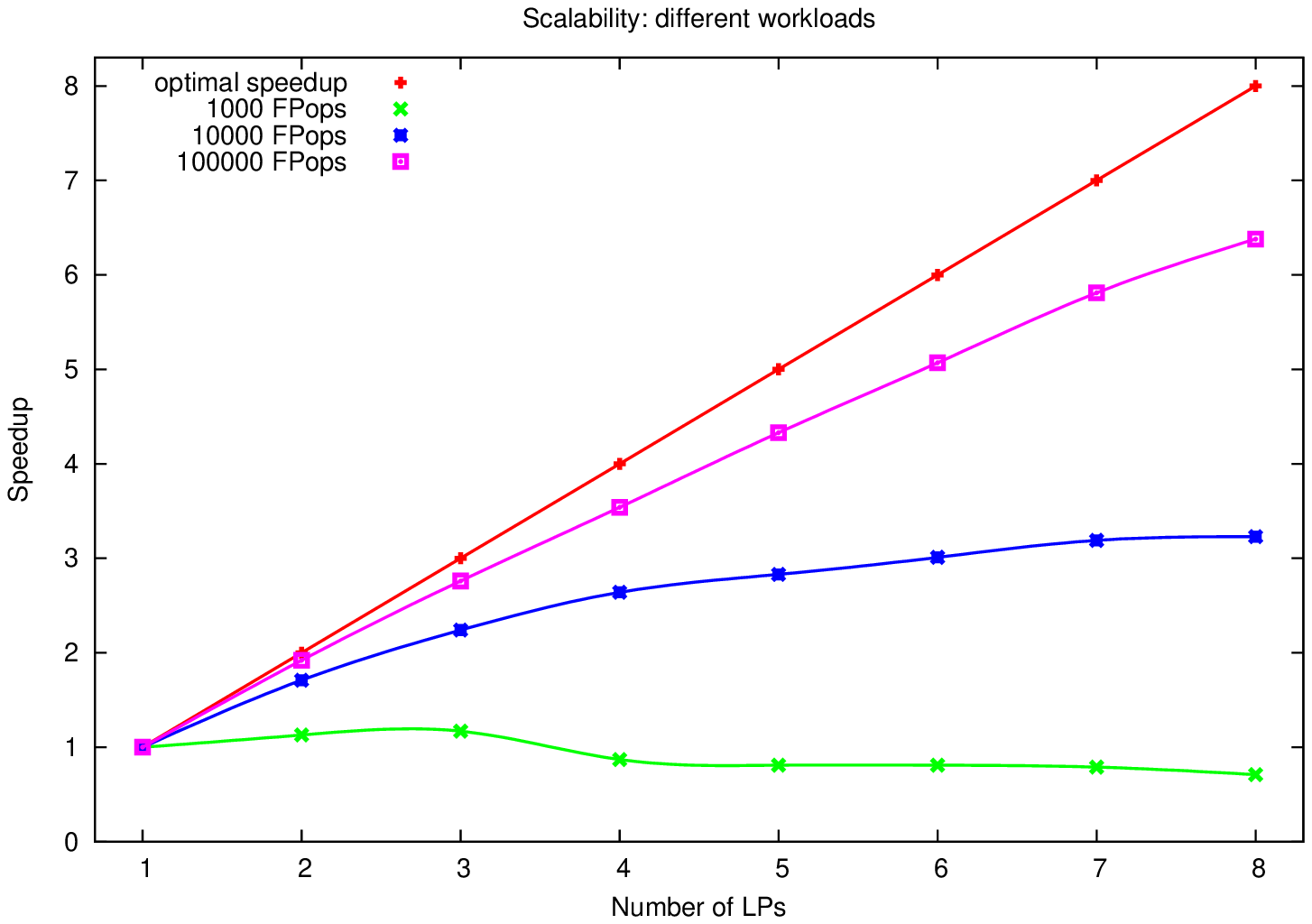}
\caption{Speedup, increasing workload per entity, different configurations}
\label{fig:workload}
\end{figure}

It turns out that running a parallel simulation gives a speedup only
when the computation load in the simulated model is enough to pay for
the extra overhead caused by communication. This means that
communication bound models are not good candidates for PADS. In
balanced systems, Go-Warp can offer a good speedup; the virtual cores
provided by HT can be used for a little increase in the performance,
but tuning the simulation setup is not straightforward (e.g.~how many
LPs and cores to use). Finally, when the model is computation bound,
going PADS is a good choice and in this case the HT can make the
difference.

\section{Conclusions}
\label{sec:conc}

In this paper we have discussed the use of multi and many-core CPUs in
the context of parallel and distributed simulation. We argued that
simulation tools should be made capable of exploiting the available
computational resources provided by modern multi-core processors in
order to improve scalability. Dealing with scalability by reducing the
size of the simulation model, or limiting the level of detail is
obviously not acceptable. In this paper we presented a parallel
simulator (Go-Warp) based on the Time Warp synchronization protocol
using the Go Programming Language. We tested Go-Warp on the PHOLD
benchmark and observed good scalability on a set of preliminary test
runs.

We plan to extend this work along many directions, first, we will work
on fine tuning the Go-Warp simulator and on a more detailed
performance evaluation. Many different setups and realistic simulation
models have to be implemented in Go and tested. Then, we will compare
the runtime performance of Go-Warp with other Time Warp
implementations based on the C/C++ language. Finally, we aim to
investigate some more radical alternatives such as the usage of
functional languages.

\section{Acknowledgments}

The authors would like to thank Pietro Ansaloni for his work on an
early version of the Go-Warp simulator.

\bibliographystyle{abbrv}
\bibliography{biblio}

\end{document}